\documentclass[a4]{PoS}
\pdfoutput=1

\usepackage{dsfont}
\usepackage{amsmath}
\usepackage{wasysym}

\title{Phase transitions in center-stabilized lattice gauge theories}

\ShortTitle{Phase transitions in center-stabilized lattice gauge theories}

\author{\speaker{Helvio Vairinhos}\\
        University of Coimbra\\
        E-mail: \email{helvio@teor.fis.uc.pt}}

\abstract{We simulate four-dimensional center-stabilized lattice Yang-Mills theories on $\mathbb{R}^3 \times S^1$ with a newly developed pseudo-heatbath algorithm. We analyze the phase structure of such theories, namely the bulk transition and the spontaneous breaking of the center symmetry associated with the compact direction.}

\FullConference{XXIX International Symposium on Lattice Field Theory \\
		 July 10 - 16 2011\\
		 Squaw Valley, Lake Tahoe, California}

\newcommand{\be}{\begin{equation}}
\newcommand{\ee}{\end{equation}}
\newcommand{\bea}{\begin{eqnarray}}
\newcommand{\eea}{\end{eqnarray}}
\renewcommand{\l}{\left}
\renewcommand{\r}{\right}
\newcommand{\Tr}{{\rm Tr}}
\newcommand{\tr}{{\rm tr}}
\renewcommand{\Re}{{\rm Re}}
\newcommand{\vecx}{{\rm\bf x}}
\newcommand{\Nfloor}{{\lfloor N/2 \rfloor}}
\newcommand{\cylinder}{{\mathbb{R}^3\times S^1}}

\begin{document}

\section{Introduction}

The idea of volume independence in large $N$ gauge theories goes back to Witten's original argument \cite{Witten79} that the path integral of $N=\infty$ Yang-Mills theory (YM) should be localized on a particular translational invariant field configuration, the master field. The translational invariance of the master field would imply the translational invariance of physical observables, 
\be
\langle {\cal O} \rangle_{L=\infty} \stackrel{N\to\infty}{=} \langle {\cal O} \rangle_{L=0}.
\ee
Consequently, spacetime degrees of freedom become spurious in the description of large $N$ physics. 

In particular, a matrix model would exist that describes the physics of pure $SU(N)$ YM on $\mathbb{R}^4$, in the $N\to\infty$ limit. The possibility of such a formulation, originally proposed in \cite{EguchiKawai}, would allow a better analytical control over YM's complicated dynamics. On the numerical side, the absence of spacetime degrees of freedom would also allow faster simulations of the theory at larger $N$. This idea of large $N$ volume reduction can be applied to other gauge theories with a similar free parameter $N$, or to the volume reduction of only a few spacetime directions (partial reduction).

However, the reduction of spacetime degrees of freedom is not freely granted. In order for a $SU(N)$ gauge theory to be independent of the volume of a particular (compact) direction, the center symmetry $Z_N$ associated with it must be intact. Technically, the expectation values of the non-contractible holonomies $\Omega_\vecx$ wrapping the reduced directions (a.k.a. reduced Polyakov loops) must be zero \cite{EguchiKawai},
\be
\langle \Tr~\Omega_\vecx \rangle \stackrel{\Omega_\vecx \mapsto z \Omega_\vecx}{=} z \langle \Tr~\Omega_\vecx \rangle = 0,~~~ z \in Z_N.
\ee

Consider four-dimensional $SU(N)$ lattice gauge theory with the standard Wilson action,
\be
S_{\rm W}(U)= -2N\lambda^{-1} \sum_x \sum_{\mu<\nu}^4 \Re\Tr ( U_{\mu,x}^{} U_{\nu,x+\hat\mu}^{} U_{\mu,x+\hat\nu}^\dag U_{\nu,x}^\dag ),
\label{eq:WilsonAction}
\ee
where $\lambda\equiv Ng^2$ is the lattice 't Hooft coupling. In this theory, large $N$ volume independence only occurs when the volume of a reduced direction is larger than a critical value, i.e. $L > L_c$, where a $Z_N$-symmetric confining phase appears \cite{KiskisNarayananNeuberger}. For $L < L_c$, the $Z_N$ symmetry breaks and the expectation values of reduced Polyakov loops become non-zero (deconfined phase). Hence large $N$ volume reduction does not hold in general. In order for the volume reduction to hold for arbitrarily small volumes, the lattice action \eqref{eq:WilsonAction} needs to be modified in a way that stabilizes the $Z_N$ symmetry.\\

\begin{figure}[t]
	\centering
	\resizebox{9cm}{!}{\includegraphics{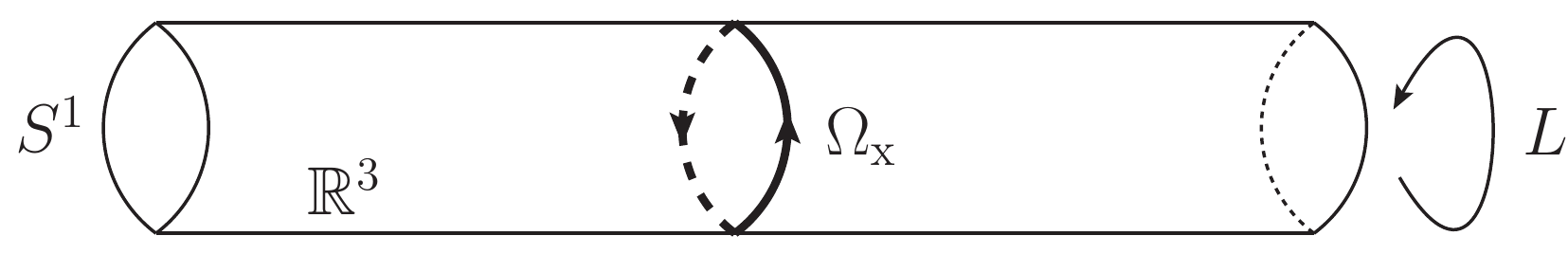}}
	\caption{Reduced Polyakov loop wrapping the compact direction of $\cylinder$.}
	\label{fig:R3xS1}
\end{figure}

Consider pure YM compactified on $\cylinder$, and let $L$ be the volume of the compact direction (Fig.\ref{fig:R3xS1}). At small $L$, the $Z_N$ symmetry is broken along the compact direction. The reason behind it resides in the fact that the effective potential of reduced Polyakov loops \cite{GrossPisarskiYaffe},
\be
V_{\rm eff}(\Omega) = -\frac{1}{L^3} \sum_{\vecx\in\mathbb{R}^3} \sum_{n=1}^\Nfloor \frac{2}{\pi^2 n^4} |\Tr~ \Omega^n_\vecx|^2
\label{eq:EffectivePotential}
\ee
is minimized at maximal traces: $\langle\Tr~ \Omega^n\rangle \neq 0$, $\forall n$. The $Z_N$ symmetry would be restored if the sign of \eqref{eq:EffectivePotential} was flipped, a situation in which $V_{\rm eff}$ would minimize at vanishing traces. Such a stabilization of the $Z_N$ symmetry can be achieved by either adding massive adjoint fermions with periodic boundary conditions along the reduced directions \cite{KovtunUnsalYaffe04}, or by adding double-trace deformations that counteract the $Z_N$-breaking character of the effective potential \cite{UnsalYaffe08},
\be
S_{\rm dYM}(U) = S_{\rm W}(U) + \frac{1}{L^3} \sum_{\vecx\in\mathbb{R}^3} \sum_{n=1}^\Nfloor a_n |\Tr~ \Omega^n_\vecx|^2,
\label{dYMAction}
\ee
known as deformed YM (dYM). For sufficiently large deformation parameters $a_n$ the $Z_N$ symmetry is restored for all $L$, making dYM theory on $\cylinder$ equivalent to pure YM on $\mathbb{R}^4$ in the $N\to\infty$ limit (where deformations become irrelevant). See also M. \"{U}nsal's talk in these Proceeedings \cite{UnsalLat11}.

In this talk we summarize the results of our non-perturbative study \cite{Vairinhos11} of the phase structure of dYM on $\cylinder$ (and related question of large $N$ volume independence), for which we performed numerical simulations on a lattice with a fully reduced direction ($L=1$).

\section{Monte Carlo algorithm}

The numerical simulation of $SU(N)$ lattice gauge theories with the standard Wilson action \eqref{eq:WilsonAction} can be performed efficiently using the Cabibbo-Marinari pseudo-heatbath algorithm \cite{CabibboMarinari}. A Metropolis algorithm could also be used, but it is typically slower, less ergodic, needs tuning for optimal acceptance rates, and has larger autocorrelation times. Pseudo-heatbath algorithms are faster and require no tuning, but they can only be applied to lattice actions that are linear with respect to each link variable. If that condition is satisfied, the probability distribution for a link $U$ reduces to
\be
\rho(U)= e^{\Re\Tr(V^\dag U)},
\ee
where $V$ is the relevant contribution from neighboring links, which must not include $U$ explicitly.

In the case of dYM, a pseudo-heatbath algorithm cannot be applied directly, since the deformation terms in \eqref{dYMAction} are not linear in the link variables. However, by introducing a sufficient number of auxiliary free scalar lattice fields $\widetilde M_i$ in the dYM partition function, and performing adequate Hubbard-Stratonovich transformations, $\widetilde M_i\mapsto M_i(\widetilde M)$, 
\bea
Z_{\rm dYM} &=& \int DU~ e^{-S_{\rm dYM}(U)} \times \underbrace{\prod_i \int d\widetilde M_i^{} d\widetilde M_i^\dag e^{-\frac{1}{2}\Tr(\widetilde M_i^{} \widetilde M_i^\dag)}}_{\rm constant} \\
&\stackrel{\widetilde M_i\mapsto M_i(\widetilde M)}{=}& \int DU~ DM^\dag DM~ e^{-\frac{1}{2}\Tr(M^\dag M)} e^{-S_{\rm W}(U)} e^{\Re\Tr( V(U,M,M^\dag)^\dag \Omega )},
\eea
we are able to linearize the dYM action \eqref{dYMAction} with respect to the link variables along the compact direction, $U\equiv\Omega_\vecx$. Consequently, we can construct a useful pseudo-heatbath algorithm for lattice dYM \cite{VairinhosMC}. This trick was inspired on a pseudo-heatbath algorithm constructed in a similar way by Fabricius and Haan for the twisted Eguchi-Kawai reduced models \cite{FabriciusHaan84}. 

The auxiliary fields $\widetilde M_i \equiv (\widetilde R_{n,\vecx}, \widetilde Q_{n,\vecx}^{(m)})$ and respective Hubbard-Stratonovich transformations that linearize \eqref{dYMAction} are given by\footnote{To we denote the normalized trace by
$\tr\equiv\frac{1}{N}\Tr$.}
\bea
\widetilde R_{n,{\rm x}} &=& \sqrt{2NL^{-3}a_n}~ \l( R_{n,{\rm x}} - \l( \Omega_{\rm x}^n - \tr\l(\Omega_{\rm x}^n\r)\mathds{1} \r) \r),~~~ 1 \leq n \leq {\Nfloor},
\\
Q^{(0)}_{n,{\rm x}} &\equiv& R_{n,{\rm x}} - \tr( R_{n,{\rm x}} ) \mathds{1},
\\
\widetilde Q^{(m)}_{n,{\rm x}} &=& \sqrt{2NL^{-3}a_n}~ ( Q^{(m)}_{n,{\rm x}} - ( Q_{n,{\rm x}}^{(m-1)} \Omega_{\rm x}^\dag + \Omega_{\rm x}^{n-m} ) )
,~~~ 1 \leq m < n \leq {\Nfloor},
\\
\widetilde Q_{n,{\rm x}}^{(m)} &=& \sqrt{2NL^{-3}a_n}~ ( Q_{n,{\rm x}}^{(m)} - \tr( R^{}_{n,{\rm x}} ) \Omega_{\rm x}^\dag ),~~~ 2 \leq n \leq m = {\Nfloor}.
\eea

For the update of the links along the compact direction, $\Omega_\vecx$, the contribution $V\equiv V_\vecx$ coming from the neighboring ``staples'' and deformations is given by
\be
V_\vecx= 2N\lambda^{-1} \sum_{\nu=1}^3 \l( \Sigma_{\vecx,\nu}^{(-)} + \Sigma_{\vecx,\nu}^{(+)} \r) + 2NL^{-3} f_\vecx, 
\ee
where $\Sigma_{\vecx,\nu}^{(-)}$ and $\Sigma_{\vecx,\nu}^{(+)}$ are the usual backward and forward ``staple'' contributions at $\vecx$, and $f_\vecx$ is the contribution coming from the deformations, encoded as a function of the auxiliary fields,
\be
  f^{}_\vecx =
  a_1 Q_{1,\vecx}^{(0)}
+ \sum_{n=2}^{\lfloor N/2 \rfloor} a_n \l( 
Q_{n,\vecx}^{(n-1)}
+ \sum_{m=1}^{n-1} Q_{n,\vecx}^{(m)\dag} Q_{n,{\rm\bf x}}^{(m-1)} 
+ \tr( R_{n,\vecx}^{} ) Q_{n,\vecx}^{(\lfloor N/2\rfloor)\dag} 
\r).
\ee

\begin{figure}[b]
	\centering
	\resizebox{7.5cm}{!}{\includegraphics{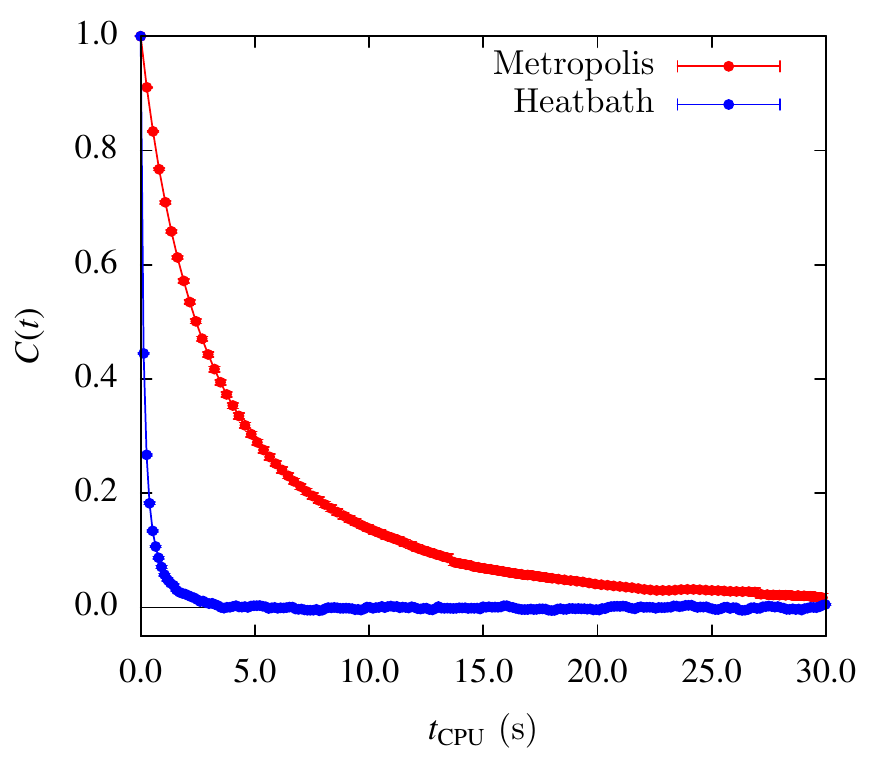}}
	\resizebox{7.5cm}{!}{\includegraphics{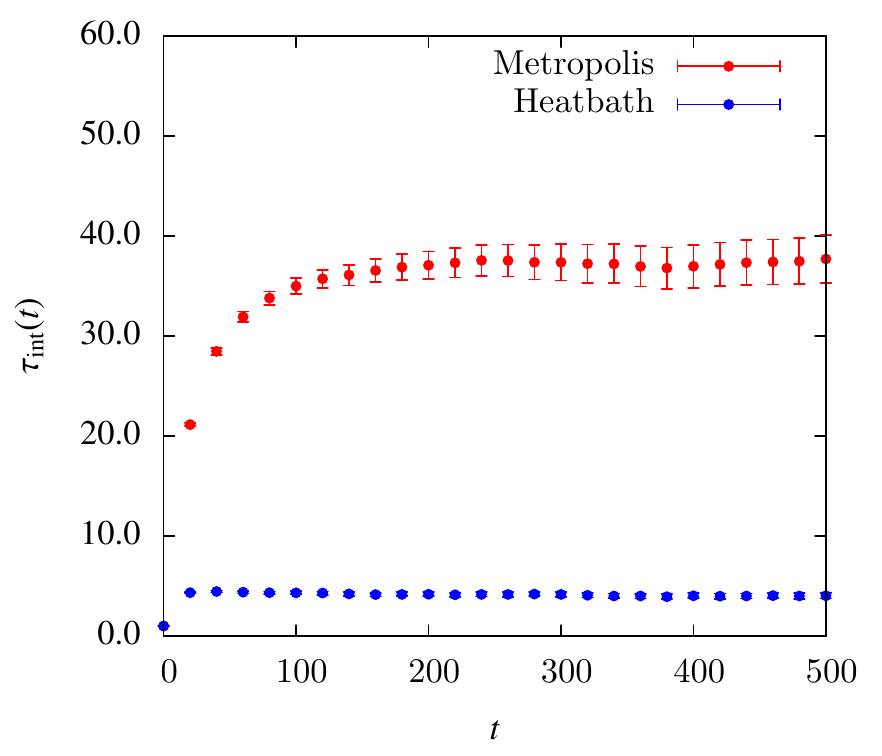}}
	\caption{Estimator of the autocorrelation function ($C$) vs. CPU time ($t_{\rm CPU}$) in simulations of $SU(5)$ dYM theory on a $10^3 1$ lattice, with $\lambda^{-1}=0.5$ and $(a_1, a_2)=(0.20, 0.05)$ (left), and the corresponding estimator of the integrated autocorrelation time ($\tau_{\rm int}$) vs. Monte Carlo time ($t$) (right). We compared a Cabibbo-Marinari-Metropolis algorithm for dYM (red) with the pseudo-heatbath algorithm described in the text (blue).}
	\label{fig:autocorrelation}
\end{figure}

We tested the pseudo-heatbath algorithm by using it in simulations of $SU(5)$ dYM on a $10^3 1$ lattice, for several couplings $\lambda^{-1}$ and deformation parameters $(a_1, a_2)$, which we compared with simulations of the same parameters using a Cabibbo-Marinari-Metropolis algorithm \cite{HasenbuschNecco}. Expectation values of the plaquettes coincide in both algorithms for all cases, which confirms the validity of the pseudo-heatbath algorithm. In the pseudo-heatbath case, however, there was a significant improvement over Metropolis in terms of equilibration and autocorrelation times (see Fig.\ref{fig:autocorrelation}).

\section{Phase structure}

\begin{figure}[t]
	\centering
	\resizebox{7.5cm}{!}{\includegraphics{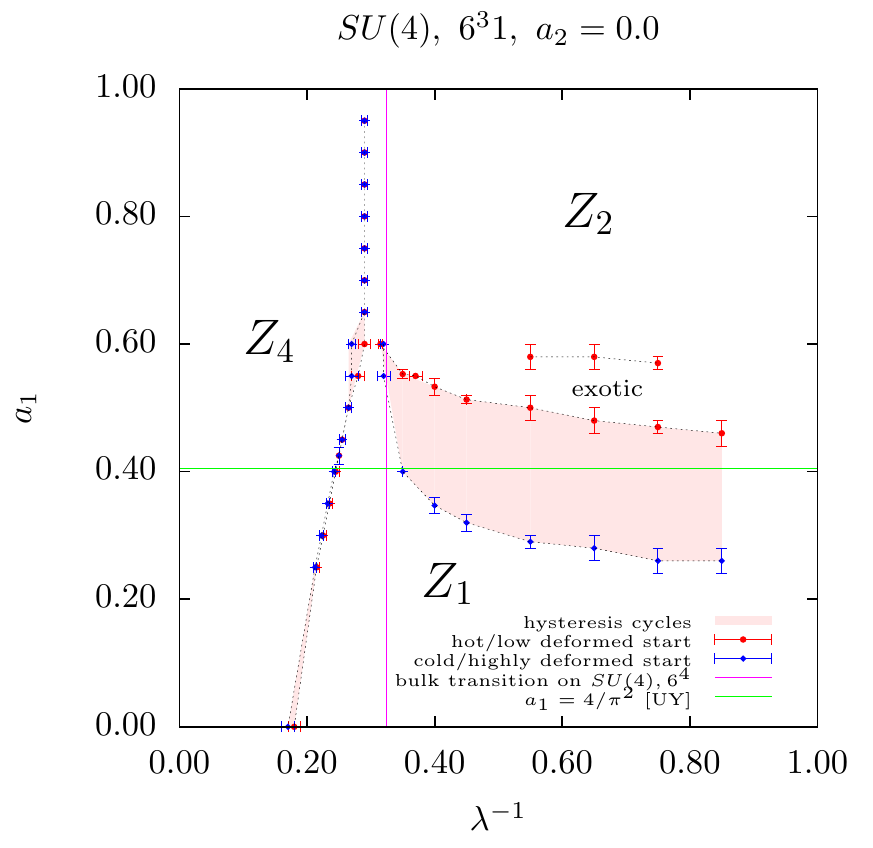}}
	\resizebox{7.5cm}{!}{\includegraphics{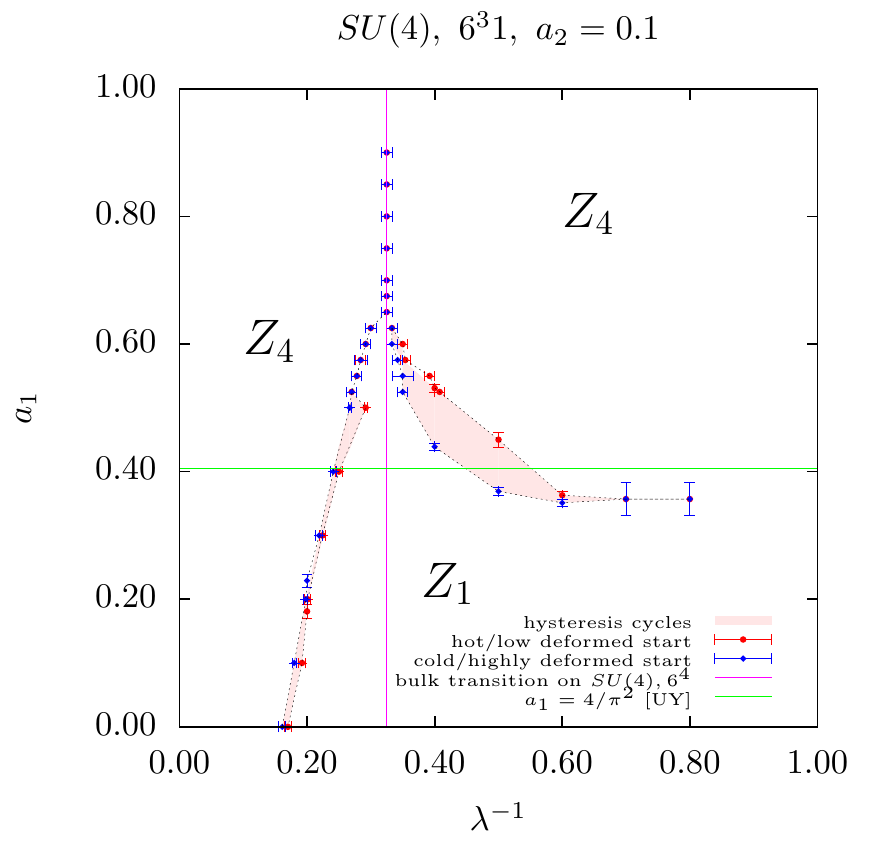}}
	\caption{Phase diagrams on the $(\lambda^{-1},a_1)$ plane of $SU(4)$ dYM on a $6^3 1$ lattice, for vanishing double-winding deformation (left) and large double-winding deformation (right).}
	\label{fig:dYM:su4}
\end{figure}

\begin{figure}[t]
	\centering
	\resizebox{7.5cm}{!}{\includegraphics{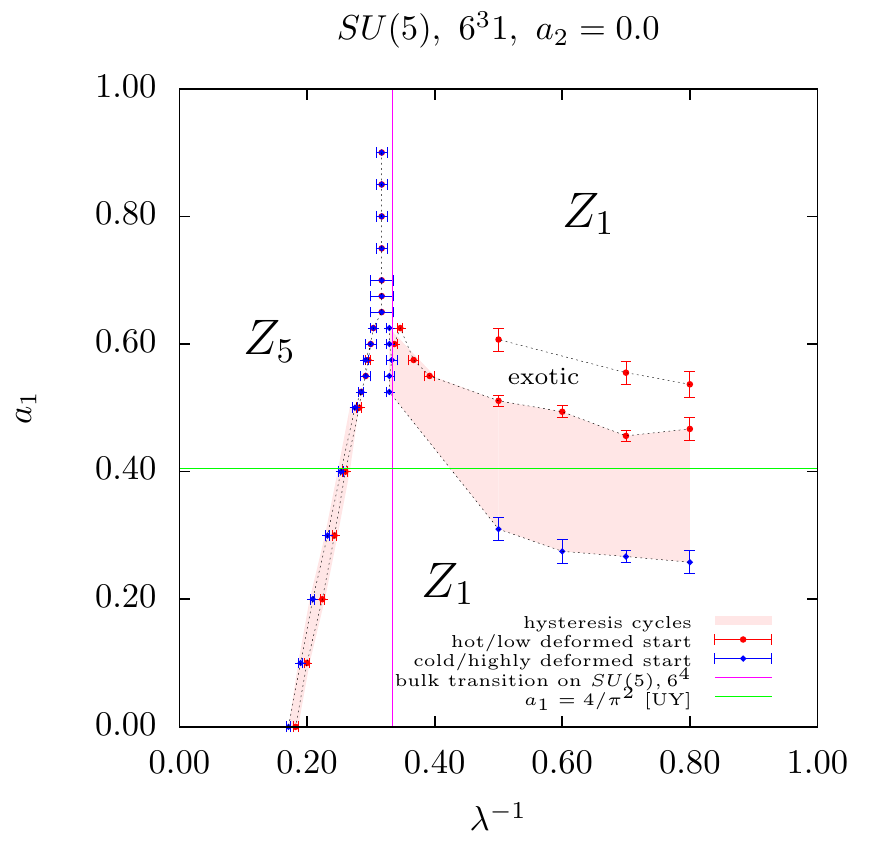}}
	\resizebox{7.5cm}{!}{\includegraphics{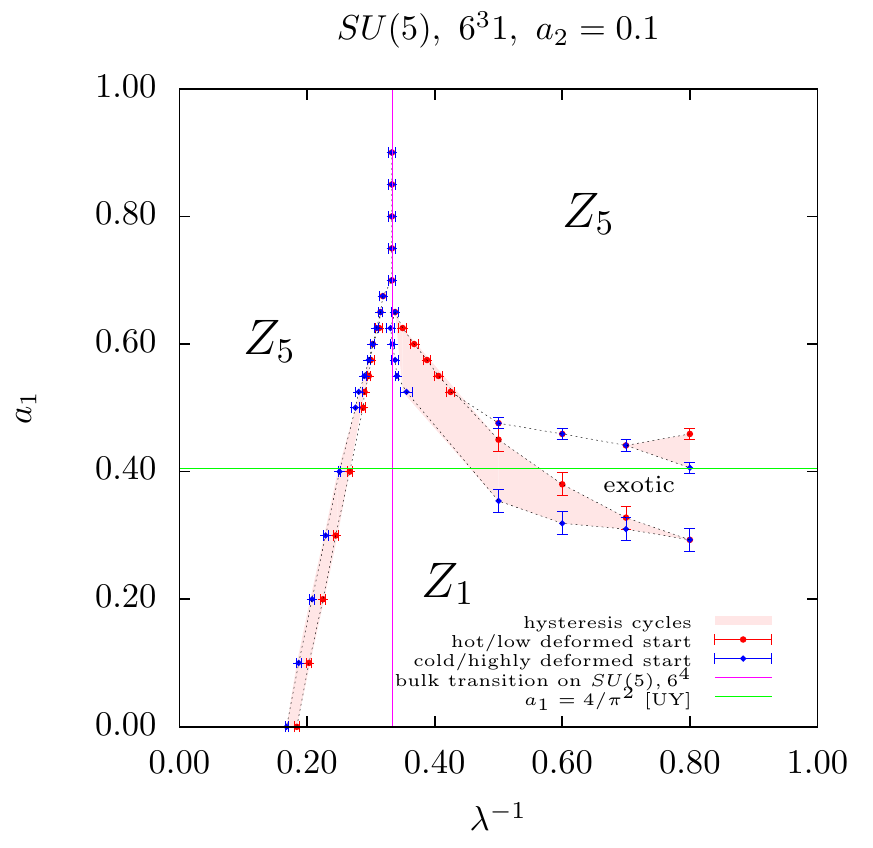}}
	\caption{Phase diagrams on the $(\lambda^{-1},a_1)$ plane of $SU(5)$ dYM on a $6^3 1$ lattice, for vanishing double-winding deformation (left) and large double-winding deformation (right).}
	\label{fig:dYM:su5}
\end{figure}

Using the pseudo-heatbath algorithm described in the previous section, we simulated $SU(N)$ dYM on a $6^3 1$ lattice, for $N=4,5$.\footnote{For these gauge groups, dYM requires a pair of independent double-trace deformations, given by single- and double-winding reduced Polyakov loops, and weighted respectively by the free parameters $a_1$ and $a_2$.} We determined the phase diagrams of these theories (Figs.\ref{fig:dYM:su4}-\ref{fig:dYM:su5}) as functions of the coupling $\lambda^{-1}$ and of the single-winding deformation parameter $a_1$. For that purpose we calculated the expectation value of reduced Polyakov loops, $\langle\tr~\Omega_\vecx^n \rangle$, which are order parameters of the $Z_N$ symmetry associated with the reduced direction, and its subgroups.

For all cases, at small $a_1$, there is a bulk transition between a $Z_N$-symmetric strongly-coupled bulk phase (left) and a $Z_N$-broken weakly-coupled deconfined phase (right). The hysteresis region (shaded area) suggests that this is a first-order transition. The breaking of the $Z_N$ symmetry in this regime of the deformation parameters indicates, as expected, that the contribution of the deformation terms in the dYM action is negligible and is not sufficient to restore the $Z_N$ symmetry at small volumes.

At large fixed $a_1$ there is a smoother bulk transition that occurs at a fixed value of the lattice coupling, $\lambda^{-1}_c$. At $\lambda^{-1} < \lambda^{-1}_c$, the phase is the same $Z_N$-symmetric strongly-coupled bulk phase. At $\lambda^{-1} > \lambda^{-1}_c$, however, the weakly-coupled phase is not always deconfining. If we consider a large value of the double-winding deformation parameter, e.g. $a_2= 0.1$, the weakly-coupled phase is $Z_N$-symmetric, i.e. confining. This indicates that the double-trace deformations are large enough to counteract the $Z_N$-breaking character of \eqref{eq:EffectivePotential}, thus preserving the $Z_N$ symmetry at all couplings.

At large $a_1$ but vanishing $a_2$, however, the $Z_N$ symmetry is partially broken to a subgroup ($Z_2$ for $N=4$, and $Z_1$ for $N=5$). This is due to the fact that $a_2$ is not large enough to preserve the subgroup $Z_p\subset Z_N$ of which $\langle\tr~\Omega_\vecx^2\rangle$ is an order parameter. Therefore, the transition $Z_N\to Z_p$ occurs.

In sum, the $Z_N$ symmetry in dYM is preserved at all couplings only when each of the deformation parameters $a_n$ is sufficiently large. In this situation, a large $N$ orbifold equivalence should hold, and dYM on $\cylinder$ should reproduce ordinary YM on $\mathbb{R}^4$ up to $O(1/N^2)$ corrections. With this in mind, we also simulated $SU(N)$ YM with the standard Wilson action on a $6^4$ lattice. We located the bulk transition for $N=4,5$ and plotted it as a vertical magenta line in Figs.\ref{fig:dYM:su4}-\ref{fig:dYM:su5}. We observe that the bulk transitions of dYM on a $6^3 1$ lattice and of YM on a $6^4$ lattice coincide when the $Z_N$ symmetry is intact in the weakly-coupled phase; but they do not coincide when the $Z_N$ symmetry is (partially) broken there. This result indicates that the expected large $N$ equivalence between dYM on $\cylinder$ and YM on $\mathbb{R}^4$ may hold quite accurately even for small $N$.

There should also be a phase transition between the regime of small deformations, $a_1 \apprle a_{1,c}$ (where the $Z_N$ symmetry is broken), and the regime of large deformations, $a_1 \apprge a_{1,c}$ (where $Z_N$ symmetry is intact and volume reduction holds). \"{U}nsal and Yaffe suggest that the critical values of the deformation parameters are given by $a_{n,c}= 4/\pi^2 n^2$ \cite{UnsalYaffe08}. Such a choice corresponds to adding $-2V_{\rm eff}$  to the standard Wilson action \eqref{eq:WilsonAction}. Perturbatively, the effective potential of dYM would be $-V_{\rm eff}$, which minimizes at vanishing traces of the the reduced Polyakov loops, $\langle\Tr~ \Omega_\vecx^n \rangle = 0$. This prediction is represented in Figs.\ref{fig:dYM:su4}-\ref{fig:dYM:su5} by an horizontal green line. 

In our simulations we observe a large hysteresis associated with the transition between small and large values of $a_1$, which suggests that it must be strongly first-order. The transition lines at asymptotic values of the lattice coupling roughly approach the value $a_{1,c}\approx 4/\pi^2$, which supports the prediction of \"{U}nsal and Yaffe.


\section{Conclusions}

We constructed an efficient Monte Carlo algorithm for $SU(N)$ dYM on $\mathbb{R}^3\times S^1$ that allows the direct use of the Cabibbo-Marinari algorithm for updates of the link variables. This algorithm has better equilibration and autocorrelation times than an optimized Metropolis algorithm for dYM.

We simulated $SU(N)$ dYM on a $6^3 1$ lattice for $N=4,5$ and mapped their phase diagrams with respect to the expectation values of (multi-winding) reduced Polyakov loops, which are order parameters of the $Z_N$ symmetry associated with the fully reduced direction (and its subgroups).

Lattice dYM possesses a rich phase structure, with phases similar to the ones discussed in \cite{OgilvieMeisingerMyers}. These include confining, deconfining and partially confining phases.

At vanishing values of the double-winding deformation parameter, $a_2$, the center symmetry breaks partially to a subgroup of $Z_N$, because $\langle\tr~\Omega_\vecx^2\rangle$ acquires a non-zero value. Only for sufficiently large values of both $a_1$ and $a_2$ is the $Z_N$ symmetry fully preserved at all couplings. 

When the $Z_N$ symmetry is intact at all couplings, lattice dYM on $\cylinder$ seems to reproduce ordinary lattice YM on $\mathbb{R}^4$. This is suggested by the fact that the critical couplings of the bulk transition in both theories coincide quite accurately.

\section*{Acknowledgments}
We are very grateful for stimulating discussions with Mike Teper, Jo\~{a}o Penedones, Masanori Hanada and Mithat \"{U}nsal. We are also very grateful for the hospitality at GGI (Florence) where part of this work was developed. Our lattice calculations were carried out on a laptop equipped with an Intel Core i7-2630QM. HV is supported by FCT (Portugal) under the grant SFRH/BPD/37949/ 2007, and under the grants PTDC/FIS/100968/2008 and CERN/FP/116383/2010 through PTQCD (Portuguese Lattice QCD Collaboration).

\end{document}